# Comment: Fisher Lecture: Dimension Reduction in Regression

**Bing Li**

This paper puts dimension reduction into the historical context of sufficiency, efficiency and principal component analysis, and opens up an avenue toward efficient dimension reduction via maximum likelihood estimation of inverse regression. I congratulate Professor Cook for this insightful and groundbreaking work. My discussion will focus on two points that explore and extend Cook's ideas. The first is about the relationship between the principal component analysis of the predictor and the regression of the response on the predictor; the second explores various ways of extending Cook's inverse regression to characterize and estimate variance components.

## 1. PCA OF $X$ AND REGRESSION OF $Y$

In his paper Professor Cook has told an intriguing and fascinating history of the opposing views regarding the relationship between the principal component analysis of $X$ and the regression of $Y$ on $X$. On the one hand, it is often the case in practice that the first few principal components of $X$ tend to have higher correlations with $Y$ than the other principal components of $X$, but on the other hand there seems no logical reason to believe that the direction along which $X$ varies the most should somehow have a relation with $Y$. In this section I ask, and attempt to answer, the following question: is it possible for the first principal component of $X$ to have higher correlation with $Y$ (than the other principal components of $X$) even if nature is "neutral" in assigning a relation between $X$ and $Y$ and "arbitrary" in assigning a covariance matrix to $X$?

*Bing Li is Professor, Department of Statistics, Pennsylvania State University, University Park, Pennsylvania 16802, USA e-mail: bing@stat.psu.edu.*

To pursue this curiosity let us consider the following situation. Let $\mathbb{R}_+^{p\times p}$ be the collection of all $p$ by $p$ positive definite matrices, and let $F$ be a distribution over $\mathbb{R}_+^{p\times p}$ that is in some sense uniform. Suppose nature randomly selects a covariance matrix $\Sigma$ according to $F$, and generates $X$ from $N(0,\Sigma)$. Furthermore, suppose that nature selects a linear relation between $X$ and $Y$ completely independently of the way it selected $\Sigma$; that is, $Y = \beta^T X + \varepsilon$, where $\beta$ is a random vector in $\mathbb{R}^p$, $\beta \perp\!\!\!\perp (\Sigma, X)$, and $\varepsilon \perp\!\!\!\perp (X, \beta, \Sigma)$ (here $\perp\!\!\!\perp$ indicates independence). Let $v_1, \ldots, v_p$ be the eigenvectors of the random matrix $\Sigma$, arranged so that their eigenvalues satisfy $\lambda(v_1) \geq \cdots \geq \lambda(v_p)$. Let $\rho_i(\beta, \Sigma)$ be the correlation coefficient between $v_i^T X$ and $Y$, conditioning on $\beta$ and $\Sigma$. Thus $\rho_1(\beta, \Sigma), \ldots, \rho_p(\beta, \Sigma)$ are random variables depending on $\beta$ and $\Sigma$. The question is: does $|\rho_1(\beta, \Sigma)|$ in any sense tend to be larger than $|\rho_2(\beta, \Sigma)|, \ldots, |\rho_p(\beta, \Sigma)|$?

To make the situation as simple as possible we take $p = 2$. We consider two ways of generating $\Sigma$ "uniformly" over $\mathbb{R}_+^{2\times 2}$. Let $\lambda_1, \lambda_2$ be i.i.d. $U(0,c)$, where $c$ is a large number, say $c = 1000$. Let $A$ be a random rotation matrix, say

$$A = \begin{pmatrix} \cos\theta & \sin\theta \\ -\sin\theta & \cos\theta \end{pmatrix},$$

where $\theta \sim U(0, 2\pi)$ and $\theta \perp\!\!\!\perp (\lambda_1, \lambda_2)$. Let

$$\Sigma = A[\mathrm{diag}(\lambda_1, \lambda_2)] A^T.$$

Intuitively, we first create a horizontal (or vertical) ellipse with arbitrary lengths of axes and then rotate it to an arbitrary angle $\theta$. Since $c$ is large this provides a reasonable approximation to a uniformly distributed $\Sigma$ over $\mathbb{R}_+^{2\times 2}$. Let $X$, $\beta$ and $Y$ be generated according to the procedure described in the last paragraph, with $\beta \sim N(0, I_p)$. For simplicity, we take $\varepsilon = 0$ because it has no bearing on the problem. We compute the probability

(1) $$P\{\rho_1(\beta,\Sigma) > \rho_2(\beta,\Sigma)\}$$

by simulation, as follows. First, generate an i.i.d. sample $(\Sigma_1,\beta_1),\ldots,(\Sigma_n,\beta_n)$. For each $(\beta_i,\Sigma_i)$, generate an i.i.d. sample $(X_{i1},Y_{i1}),\ldots,(X_{im},Y_{im})$. Using this sample we estimate $\rho_1(\beta_i,\Sigma_i)$ and $\rho_2(\beta_i,\Sigma_i)$ by the method of moments. Denote these estimates by $\hat\rho_{i1},\hat\rho_{i2}$. Finally, we use the relative frequency of the cases $\hat\rho_{i1} > \hat\rho_{i2}$ among the sample $(\Sigma_1,\beta_1),\ldots,(\Sigma_n,\beta_n)$ to estimate the probability (1). Taking $m = n = 200$, this probability is estimated to be 0.65, larger than one half.

An alternative way of generating uniform $\Sigma$ is as follows. Generate $(\lambda_1,\lambda_2)$ as before. Then, generate $\alpha$ from $U(-\sqrt{\lambda_1\lambda_2},\sqrt{\lambda_1\lambda_2})$, and define

$$\Sigma = \begin{pmatrix} \lambda_1 & \alpha \\ \alpha & \lambda_2 \end{pmatrix}.$$

Under this alternative scheme we recalculated the probability (1) to be 0.735, again larger than one half.

I have tried several distributions for $\beta$ and values of $c$, and the probability (1) is invariably greater than one half. Thus it seems reasonable to make the following conjecture [we will abbreviate the random variable $\rho_i(\beta,\Sigma)$ by $\rho_i$].

CONJECTURE 1.1. *Suppose $\Sigma$ is a random matrix uniformly distributed over $\mathbb{R}_+^{p\times p}$, and suppose $X \sim N(0,\Sigma)$ and $Y = \beta^T X + \varepsilon$ with $\beta \perp\!\!\!\perp (X,\Sigma)$, $\varepsilon \perp\!\!\!\perp (X,\beta,\Sigma)$ and $\varepsilon \sim N(0,\sigma^2)$. Then, for any $i \in \{2,\ldots,p\}$,*

(2)
$$\begin{aligned} &P(|\rho_1| = \max\{|\rho_1|,\ldots,|\rho_p|\}) \\ &\quad > P(|\rho_i| = \max\{|\rho_1|,\ldots,|\rho_p|\}).\end{aligned}$$

This conjecture, if true, does seem to suggest that, if nature selects an arbitrary covariance matrix for $X$ and an arbitrary linear relation between $X$ and $Y$, then the first principal component of $X$ tends to have the largest correlation with $Y$ among all principal components of $X$.

To see why this conjecture should hold, imagine the extreme case where support of $X$ is concentrated on a line. In this case the only way for $Y$ to be correlated with $X$ is to be correlated with its first principal component. Intuitively, this tendency should still hold when the distribution of $X$ is not concentrated on a line but has elongated elliptical contours. Now, if nature draws $\Sigma$ from a uniform distribution, there is a nonzero probability that the distribution of $X$ has elongated contours, in which case the projection of $X$ onto the longest axis of the ellipsoid tends to have largest correlation with $Y$ (among its projections onto other axes), even if $\beta$ is drawn independently from $\Sigma$. In the cases where $X$ does not have elongated contours, $|\rho_1|$ would not stand out as the largest, but then neither would the other $\rho_i$'s. Thus, on average, something like (2) should hold.

The above example also shows that the tendency (2) is a modest one. When $p = 2$ the probability (1) is around $65\% \sim 75\%$, only modestly larger than $50\%$. Similarly, when $p$ is larger than 2 I do not expect this probability to be drastically larger than $1/p$ [which is the probability in (1) when $\rho_1,\ldots,\rho_p$ are symmetric]. Thus there should still be a substantial gain in performing dimension reduction of $X$ in reference to $Y$.

## 2. INVERSE REGRESSION FOR PRINCIPAL VARIANCE COMPONENT

What is interesting about Cook's inverse regression model [model (2) in Cook's paper] is that the parameter $\Gamma$ automatically provides sufficient dimension reduction for the forward model, in the sense that $Y \perp\!\!\!\perp X | \Gamma^T X$. The same idea can be used to construct an inverse regression model where the conditional variance $\text{var}(X|y)$, rather than the conditional mean $E(X|y)$, depends on $y$. Such models would be useful in the classification problems where the several groups involved differ in their dispersions but not so much in their locations. See, for example, Cook and Yin (2001) for a breast cancer data set whose behavior roughly fits this description.

Consider the inverse regression model

(3) $$X = \sigma^2(\Gamma \nu_y \Gamma^T + I_p)\varepsilon,$$

where $\nu_{(\cdot)} : \Omega_Y \to \mathbb{R}^{d\times d}$, $d < p$ and $\Gamma$ is a $p \times d$ semiorthogonal matrix.

THEOREM 2.1. *If $Y \perp\!\!\!\perp \varepsilon$ and if model (3) holds, then $Y \perp\!\!\!\perp X | \Gamma^T X$.*

PROOF. Let $\Gamma_0$ be a $p \times (p-d)$ semiorthogonal matrix such that $\Gamma_0^T \Gamma = 0$. Relation (3) implies the equalities

(4)
$$\begin{aligned} \Gamma^T X &= \sigma^2(\nu_y + I_d)\Gamma^T \varepsilon, \\ \Gamma_0^T X &= \sigma^2 \Gamma_0^T \varepsilon.\end{aligned}$$

By the assumption $\varepsilon \perp\!\!\!\perp Y$, conditioning on $Y$, $\Gamma_0^T X$ and $\Gamma^T X$ are multivariate normal with conditional covariance

$$\text{cov}(\Gamma^T X, \Gamma_0^T X | Y = y) = (\nu_y + I_d)\Gamma^T \Gamma_0 = 0.$$

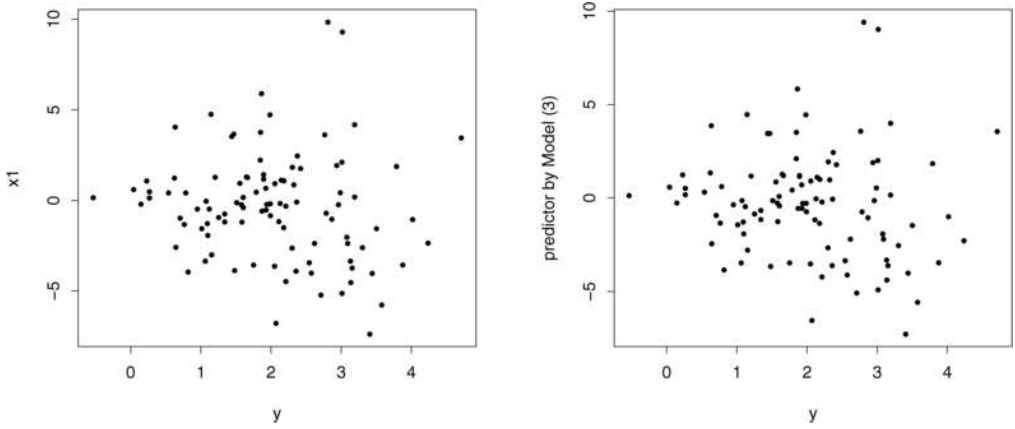

FIG. 1. *Dimension reduction via model (3). Left panel: $X_1$ versus $Y$; right panel: $\hat{\Gamma}^T X$ versus $Y$.*

Hence $\Gamma^T X \perp\!\!\!\perp \Gamma_0^T X | Y$. Meanwhile, from the second equality in (4) we see that $\Gamma_0^T X \perp\!\!\!\perp Y$. Hence $Y \perp\!\!\!\perp X | \Gamma^T X$. □

To see how this model can be used in practice, we consider the following example as an illustration.

EXAMPLE 2.1. We take $p=3$ and $d=1$, $\nu_y = |y|$ and $\Gamma^T = (1,0,0)$. Assume $Y \sim N(2,1)$, $Y \perp\!\!\!\perp \varepsilon$ and $\varepsilon \sim N(0, I_p)$. Thus we have the inverse regression model

$$X = \begin{pmatrix} 1+|y| & 0 & 0 \\ 0 & 1 & 0 \\ 0 & 0 & 1 \end{pmatrix} \varepsilon.$$

We generate $(X_1, Y_1), \ldots, (X_n, Y_n)$, where $n = 100$, from model (3), and estimate $\Gamma$ by a numerical maximization of the likelihood, which gives

$$\hat{\Gamma}^T = (0.964, 0.047, 0.068).$$

Figure 1 presents the scatterplots of $X_1$ versus $Y$ (left panel) and $\hat{\Gamma}^T X$ versus $Y$ (right panel). We can see that they are very much in agreement.

We can further generalize model (3) to accommodate the situations where both the location and the dispersion in the inverse regression model depend on $y$, by combining model (3) above and model (2) in Cook's paper, as follows:

(5) $\quad X = \mu + \Gamma_1 \nu_y + \sigma^2 (\Gamma_2 \tau_y \Gamma_2^T + I_p) \varepsilon,$

where $\varepsilon \sim N(0, I_p)$, $\varepsilon \perp\!\!\!\perp Y$, $\Gamma_1 \in \mathbb{R}^{p \times d_1}$ and $\Gamma_2 \in \mathbb{R}^{p \times d_2}$, with $d_1 + d_2 < p$, $\nu_{(\cdot)} : \Omega_Y \to \mathbb{R}^{d_1}$ and $\tau_{(\cdot)} : \Omega_Y \to \mathbb{R}^{d_2 \times d_2}$. Here, for convenience we again assume that $\Gamma_1$ and $\Gamma_2$ are semiorthogonal matrices. Note that the column spaces of $\Gamma_1$ and $\Gamma_2$ may or may not be the same. Similarly to model (2) in Cook's paper and model (3) above, relation (5) provides automatically a sufficient dimension reduction of $X$.

THEOREM 2.2. *If model (5) holds, then $Y \perp\!\!\!\perp X | (\Gamma_1^T X, \Gamma_2^T X)$.*

PROOF. Let $\Gamma = (\Gamma_1, \Gamma_2)$, and let $\Gamma_0$ be a matrix such that the matrix $(\Gamma, \Gamma_0)$ has full row rank and $\Gamma_0^T \Gamma = 0$. Multiply both sides of equality (5) on the left by $\Gamma^T$ and $\Gamma_0^T$, respectively, to obtain

$$\Gamma^T X = \Gamma^T \mu + \Gamma^T \Gamma_1 \nu_y + \sigma^2 \Gamma^T (\Gamma_2 \tau_y \Gamma_2^T + I_p) \varepsilon,$$
$$\Gamma_0^T X = \Gamma_0^T \mu + \sigma^2 \Gamma_0^T \varepsilon.$$

Following the same argument as in the proof of Theorem 2.1, we see that $\Gamma^T X \perp\!\!\!\perp \Gamma_0^T X | Y$ and $\Gamma_0^T X \perp\!\!\!\perp Y$, from which it follows that $X \perp\!\!\!\perp Y | \Gamma^T X$. □

The next example illustrates the use of model (5), which has both a location and a dispersion component in the inverse regression.

EXAMPLE 2.2. We take $p = 3$ and $d = 1$. Assume $Y \sim N(3,1)$, $Y \perp\!\!\!\perp \varepsilon$ and $\varepsilon \sim N(0, I_p)$. Consider the inverse regression model

$$X = \begin{pmatrix} 5y \\ 0 \\ 0 \end{pmatrix} + \begin{pmatrix} 1+|y| & 0 & 0 \\ 0 & 1 & 0 \\ 0 & 0 & 1 \end{pmatrix} \varepsilon.$$

This is a special case of model (5) with $\Gamma_1 = \Gamma_2 = \Gamma$. As in Example 2.1, we generate $n = 100$ pairs of observations from this model and maximize the likelihood numerically, which gives

$$\hat{\Gamma}^T = (1.969, 0.052, 0.010).$$

We see that this estimate is more accurate than that in Example 2.1 (the contrast between the first component and the last two components is greater). This is because it uses the additional information provided by the location term. The comparison of the scatterplots of $X_1$ versus $Y$ and $\hat{\Gamma}^T X$ versus $Y$ is given in Figure 2.

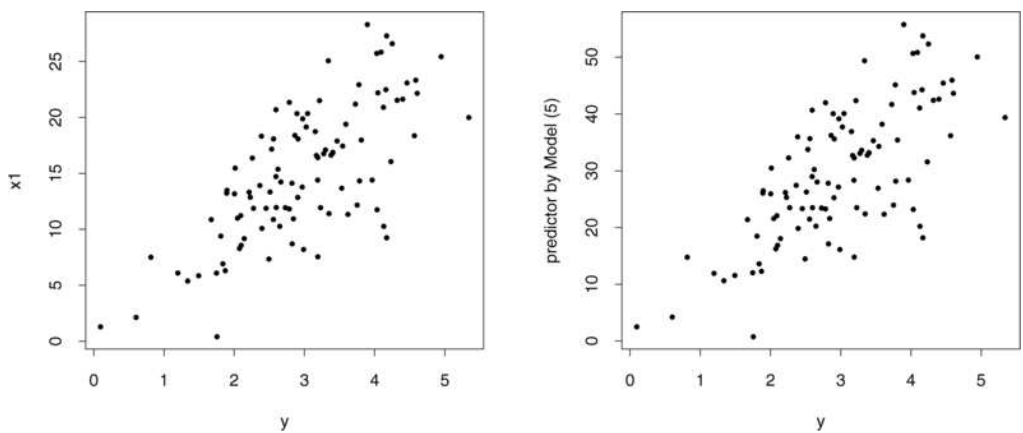

FIG. 2. *Dimension reduction via model (5). Left panel: $X_1$ versus $Y$; right panel: $\hat{\Gamma}^T X$ versus $Y$.*